\documentclass[a4paper]{article}

\begin{document}

\title{Thoughts About a Conceptual Framework for Relativistic Gravity
\thanks{To be published, without the abstract and with small editorial changes, in {\em Einstein and the Changing Worldviews of Physics} ({\em Einstein Studies}, vol.\ 12). ed C Lehner,  J Renn, M Schemmel. Boston: Birkh\"auser (2011). Based on a talk delivered at the Seventh International Conference on the History of General Relativity, Tenerife, Canary Islands, March 2005.}}

\author{Bernard F Schutz\\[3mm]
Max Planck Institute for Gravitational Physics \\(Albert Einstein Institute), 14476 Potsdam/Golm, Germany, \\and \\Department of Physics and Astronomy,\\ Cardiff University, Cardiff, Wales, UK.\\ {\tt Bernard.Schutz@aei.mpg.de}}
\date{}
\maketitle

\abstract{I consider the isolation of general relativity research from the rest of theoretical physics during the 1930s-1950s, and the subsequent reinvigoration of the field. I suggest that the main reason for the isolation was that relativists of the time did not develop heuristic concepts about the physics of the theory with which they could communicate with other physicists, and that the revival happened when they began to develop such concepts. A powerful heuristic today is the concept of a black hole, which is a robust and stable component of many astronomical systems. During the 1930s relativists could only offer the ``Schwarzschild singularity''. I argue that the change occurred at least partly because key theoretical physicists schooled in quantum theory entered relativity research and began to approach problematic issues by asking questions about observable effects and the outcomes of thought experiments. The result was the development of a physical intuition about such things as black holes, which could then be communicated to non-specialists. Only then was it possible to integrate general relativity fully into the rest of physics.}

\bigskip\noindent{\large\bf 1. Introduction}\\[2mm]
Mine is one of several talks at this meeting that consider the revival of relativity and its integration into the mainstream of physics, beginning in the 1950s. Ted Newman has described the physics problems that created confusion during the slow period 1930--1950, and how eventually a new generation of young physicists pulled the theory out of its mire. Silvio Bergia has emphasized the changes of thinking that were required, and the importance of the physical insight and especially the geometrical perspective that John Wheeler, among others, brought to the subject. I want to focus on the gulf that opened up during the slow period between relativists and the rest of what I will call mainstream theoretical physics. This gulf is important not just for the negative influence it exerted on the development of relativity. It also has much to teach us about what physicists expect from a theory of physics, and especially about the role of heuristic concepts in physicists' communication with one another.

My thesis is that general relativity, despite its essential {\em mathematical} completeness in 1916, did not become a complete theory of {\em  physics} until the 1970s. In order to understand this period, scholars of relativity need to look, not just at progress in understanding the mathematical theory, but at the slow development of heuristic concepts that were needed to enable relativists to talk to other physicists in a common language.

Today we have a fairly secure set of heuristic concepts: for example, we know what a black hole is, we know what gravitational waves do, we know how gravitational lenses work. These concepts -- black holes, gravitational waves, gravitational lenses -- have gained a kind of concrete physical reality, even though if you take them apart they are just ideas that rest ultimately on rather complex (and usually approximate) solutions of Einstein's field equations. Very importantly, they are concepts that relativists can communicate to nonrelativists who may need them (astronomers, experimental physicists, historians, the general public) without needing to pass on all their mathematical underpinnings.

The key accomplishment of the generation of physicists who revived relativity is that they created a wide range of useful concepts like these out of the confusions that plagued the previous generation. This took a huge amount of work, but the work was not done at random. Rather, a handful of creative and senior physicists, many of whom came to relativity from other branches of physics, very deliberately shaped the directions of research toward developing these paradigmatic concepts, thereby adding the physics to the mathematical skeleton of the theory. In my view, the absence of such a vision of how to make relativity into a working theory of physics was what, in the dark period, led to the increasing isolation of relativity from the mainstream.

\bigskip\noindent{\large\bf 2. The Gulf of Relativity}\\[2mm]
The gulf between mainstream physics and relativity between 1930 and 1960 is remarkable for how huge it was (Eisenstaedt 1989, 2006). Rarely has an important sub-field of physics enjoyed such a poor reputation. Very few physicists moved back and forth across the gulf or even made an effort to communicate across the divide.

Equally remarkable has been the subsequent huge turn-around. Gravitational physics is mainstream physics today. Massive amounts of money fund gravitational wave experiments; the holy grail of theoretical particle physics is to unify the nuclear and electromagnetic forces with gravity; a course in general relativity is standard for physics graduate students.  A few short anecdotes serve to illustrate the depths to which relativity sank and the heights to which it has subsequently risen.

{\em Anecdote 1.}  The Nobel-Prize-winning astrophysicist Subrahmanyan Chandrasekhar kept a remarkable scientific diary, in which at the end of each year he summarized his scientific work and decisions of that year. Shortly after Chandra's death in 1995, Norman Lebovitz (private communication) showed me some of the entries. Very interestingly, Chandra writes that, during the 1930s, he considered starting to do research in relativity, in order to explore what would happen to a compact star that exceeded the maximum white-dwarf mass that Chandra himself had recently established.  He consulted other physicists, who strongly advised him against doing this.  General relativity, one told him, had proved to be a ``graveyard of many theoretical astronomers''.  Chandra particularly mentions that Niels Bohr discouraged him from making a move into relativity. (Considering Bohr's exchanges with Einstein on the interpretation of quantum mechanics, this is a tantalizing remark!) Chandra's reputation and career were by no means secure in the 1930s, and so he looked (very productively) elsewhere for research problems. It was not until after 1960 that he felt confident enough of his reputation that he finally indulged his long-postponed wish to work on general relativity. [Kip Thorne, one of the dominant figures in modern relativity research, reports (Thorne 1994) that he had similarly negative advice when he was contemplating doing graduate work in relativity in the early 1960s.]

{\em Anecdote 2.} It is worth looking here at Richard Feynman's famous reaction (in a letter to his wife) to the 1962 Warsaw relativity meeting (Feynman 1988):
\begin{quote}
I am not getting anything out of the meeting. I am learning nothing. Because there are no experiments this field is not an active one, so few of the best men are doing work in it. The result is that there are hosts of dopes here and it is not good for my blood pressure: such inane things are said and seriously discussed that I get into arguments outside the formal sessions (say, at lunch) whenever anyone asks me a question or starts to tell me about his ``work.'' The ``work'' is always: (1) completely un-understandable, (2) vague and indefinite, (3) something correct that is obvious and self-evident, but worked out by a long and difficult analysis, and presented as an important discovery, or (4) a claim based on the stupidity of the author that some obvious and correct fact, accepted and checked for years, is, in fact, false (these are the worst: no argument will convince the idiot), (5) an attempt to do something probably impossible, but certainly of no utility, which, it is finally revealed at the end, fails, or (6) just plain wrong. There is a great deal of ``activity in the field'' these days, but this ``activity'' is mainly in showing that the previous ``activity'' of somebody else resulted in an error or in nothing useful or in something promising. It is like a lot of worms trying to get out of a bottle by crawling all over each other. It is not that the subject is hard; it is that the good men are occupied elsewhere. Remind me not to come to any more gravity conferences!
\end{quote}

This is, of course, typical Feynman hyperbole. We know that at that meeting (Infeld 1964) a core group of relativists was already coming to grips with issues like energy, black holes, and the reality of gravitational waves.  And ironically it took place just a year before the first Texas Symposium in Relativistic Astrophysics (Robinson, et al, 1965), which is often regarded as the moment that relativity began to have real interest to astrophysicists.  Nevertheless Feynman's remarks show why it would still be another couple of decades before mainstream theoretical physics would completely drop its prejudices against the relativity community. The two sides were not communicating.

{\em Anecdote 3.} I vividly remember my own personal experiences as a young relativist. In the 1970s if I mentioned black holes to an astronomer, the best I could usually hope for was a patronizing smile. And this was after the discovery of what we now know was the first black hole in a binary system, Cyg X-1, by the Uhuru satellite (which led to the award of the 2002 Nobel Prize to Riccardo Giacconi). Later, during the 1980s, when I moved into gravitational wave detection, many astronomers told me I was throwing my career away. And they were the sympathetic ones; others just saw me as a misguided threat to their own research funding!

{\em Anecdote 4.} If the low point of relativity was very low, the current high point is indeed very high. Nothing illustrates the dramatic nature of this turn-around better than money.  By 2020 at least 3-4 billion dollars will have been invested by a dozen national and international scientific organizations in building gravitational wave detectors on the ground and in space.  Most of this money has already been committed, at least in a planning sense, and this has all happened even before the first direct detection of a gravitational wave!

Where has today's immense faith in general relativity come from? How did relativity establish such strong credentials after being in such disrepute? It seems to me that to answer this question we need to do more than simply catalog the details of what happened in relativity and astrophysics to get us where we are today. We have to understand how physicists judge the credibility of other physicists. The tortured development of gravitational physics is a good case study of how physicists decide that other people are really doing physics, even though they may not understand the mathematical and technical details.

\bigskip\noindent{\large\bf 3. Heuristics in General Relativity}\\[2mm]
I won't attempt to give anything like a complete set of answers to the questions I have just posed, but I think a key to answering them lies in the fact that physicists have a characteristic way of thinking, which they call physical intuition. Physicists think in terms of models, of heuristic concepts that they connect up using this physical intuition. Physicists' models must of course be founded on the mathematical expression of a theory, but physicists are typically not happy if all they have are mathematical links between their models. They want concepts that enable them to understand essential parts of theories, even if they have not developed a facility with the mathematics of those theories. They need to have models they can exchange with physicists in other specialties, which allow those physicists to work with the concepts without being expert in their underlying theory.

I will illustrate the conceptual changes in relativity between the ``dark ages'' and the modern era by considering two key issues that were also listed by Ted Newman in his talk as key problems that were not solved during the dark years of relativity. The first is the meaning of the Schwarzschild solution. In the 1930s people talked about the ``Schwarzschild singularity'' (by which they meant the horizon, not the crunch at the center). Today we use the term ``black hole''.  There is a world of difference between the ideas behind these different terminologies.  If you think you have a singularity then you can't use it in a physical model. You don't know how to include such an object in a physical system, either as the outcome of gravitational collapse or as an object that might affect other objects with its gravitational field. On the other hand, the term ``black hole'' is a shorthand description of a real object, one which you can confidently include in models for some physical systems: as the constituent of a model for an X-ray binary system, for example, or as a gravitating center in the middle of a galaxy. In the first case you are paralyzed by incomprehension.  In the second you can hide away all the nonlinear general relativity, if you wish, and treat the object as just another member of the vast zoo of objects that makes up our fascinating universe.

My second example is gravitational radiation. In the low period, people worried about the reality of the radiation itself.  Doubting that waves could remove energy from sources and/or deposit it in detectors, relativists were unable to draw the clear parallels with electromagnetic radiation that would have emphasized the natural place that general relativity has in theoretical physics. By resolving these issues, relativists were finally in the position by 1980 to take advantage of the discovery of the Hulse-Taylor binary pulsar to show that observations supported the dynamical sector of Einstein's theory. Not all the mathematical problems associated with gravitational waves are yet solved even today, but the field has enough confidence in its approximation methods and its control over the remaining outstanding issues that it has been able to develop a thoroughly convincing physical picture of gravitational waves.

\bigskip\noindent{\large\bf 4. Why Did the Gulf Drift Open?}\\[2mm]
So why did relativists find themselves excluded from the rest of theoretical physics in the 1930s to 1950s? Apart from a few notable exceptions, such as J.\ Robert Oppenheimer and Lev Landau, hardly anyone worked in relativity and other areas of theoretical physics between 1930 and 1960. And Oppenheimer and Landau were largely ignored by relativists (Thorne 1994). Let me list some explanations that are often offered and indicate why I don't find them adequate.
\begin{enumerate}
\item General relativity is mathematically very difficult.  The combination of nonlinearity and coordinate freedom made it difficult to make definite statements.  This certainly underlay the problem that relativists had, and it explains why progress on understanding the theory was slow. But it does not explain the low regard that ``real'' physicists had for relativists.  Indeed, one might have expected them to have gained respect from the rest of physics for making even small progress with such a difficult theory.
\item As Feynman remarked, there was little experimental data.  This meant that progress relied especially strongly on the ability to ask and resolve the right kinds of theoretical questions. But one might have expected the field to have exploited the few observational hints that did exist. Chandrasekhar's upper limit on the mass of white dwarfs, coupled with Fritz Zwicky's suggestion that supernova explosions led to neutron stars, were a clear invitation to explore gravitational collapse and the Schwarzschild solution.  But only Oppenheimer and Landau seem to have found this interesting. Importantly, they were physicists who approached relativity from outside, from the point of view of the mainstream theoretical community. Moreover, it is significant that the revival of relativity started during the 1950s without the stimulus of any new experimental or observational data. So, while an abundance of data would certainly have driven the field in the right direction had it been available, I don't think that its absence explains why the field slipped into such a low state.
\item Relativity had to compete with quantum theory for good people.  As Feynman says, ``few of the best men are doing work in it''. The competition was certainly there, but I don't believe that physics was that compartmentalized in the 1930s to 1950s. Leading quantum theorists had a deep interest in general relativity; Wolfgang Pauli wrote a beautiful textbook on it. The theory was widely regarded as the supreme achievement of 20th century theoretical physics. One would think that if relativists had made their own work interesting to mainstream physicists then they would not have worked in such isolation.  There might have been many more Oppenheimers and Landaus crossing the gulf if the relativity community had welcomed them and worked with them, or even been able to communicate with them.
\item The Second World War got in the way. There is no doubt that this seriously retarded research, removing young people from research and inhibiting international scientific communication. The cold war afterwards did not help. Nuclear physics had proved so useful to the military that it (including particle physics) was well-funded after the war, whereas relativity fell into a theoretical backwater. But I am not convinced that this should have caused relativists to lose their way. Attacking the key problems of this period did not require a lot of money. A small field can still earn the respect of the majority. And the relativity community seems to have suffered less than other fields from the divisions of the cold war.  It seems clear to me that, once the revival started, it went significantly more rapidly because of the relatively free intellectual interchange between Western and Soviet-bloc scientists working in relativity.\footnote{The International Society for General Relativity and Gravitation (known as the GRG Society), which is today the main professional society for relativists worldwide, is one of the few societies adhering directly to IUPAP which has individual scientists as members, not national organizations. During the cold war this structure enabled it and its predecessor (the International Committee for General Relativity and Gravitation) to organize relatively apolitical meetings that scientists from both sides of the Iron Curtain attended. An example was the famous meeting in Warsaw that Feynman criticised.}
\end{enumerate}

I believe that the gulf opened between relativity and mainstream physics, not directly because of the problems listed above, but because the relativity community's response to at least the first two problems was to ask the wrong questions.  For example, one of the serious mathematical challenges that they faced was coordinate freedom.  Ted Newman in his talk at this meeting cataloged the way the community clearly missed opportunities to understand that the so-called Schwarzschild singularity is just a coordinate effect. To us today this episode is baffling.  Relativists do not seem to have understood the importance of controlling the effects of coordinates on their results, despite Einstein's emphasis that the physics should be coordinate-invariant. They even had coordinate systems available at that time (from the work of Sir Arthur Eddington and Georges Lema\^{i}tre) that went across the horizon in a non-singular way.

In the same years, quantum physicists were (at times painfully) revolutionizing their physical thinking, agreeing that they should only concern themselves with the results of measurements, which they called observables, and that they should not try to create physical models for what can't be measured, such as the ``paths'' of quantum particles. Special relativity already had a similar tradition, going back to Einstein's gedanken experiments, which were designed to focus attention on the outcome of experimental measurements rather than phrase the predictions of special relativity in terms of observer-dependent notions of time and space.  Yet this trend did not seem to influence research in general relativity in the period 1930--1950 as much as it should have.

Ted Newman also mentioned another example, the deep confusion over the concept of energy in space-times containing gravitational waves. The resolution of this issue only began when Hermann Bondi and his successors, who included Roger Penrose and Ted himself, discovered how to treat radiated energy far from its source. It is easy to understand why relativists felt that they needed to clarify the idea of energy: energy is one of the key heuristics of mainstream physics. However, I confess that I don't understand why relativists allowed the genuine difficulties of defining gravitational wave energy to stop their developing a physical understanding of gravitational waves themselves. It appears that, because it was difficult to define the energy of a radiating system or to localize the energy carried by waves, relativists during this period were unable to develop any kind of useful physical model for gravitational waves.

We know today that it is perfectly possible to describe the generation of gravitational waves and their action on a simple detector without once referring to energy; the quadruple formula for the generation of the waves and the geodesic equation for their action on a simple detector are all one needs, and these tools were available from 1918.  It is also possible to show that gravitational waves certainly deposit energy in some kinds of detectors, without having a full global energy conservation law.  Indeed, Feynman at the earlier relativity meeting in Chapel Hill in 1957 (Bergmann 1957) presented a simple argument to show how a gravitational wave would heat a detector that has internal friction.  The argument is so direct that I used a version of it myself in my undergraduate-level relativity textbook (Schutz 2009), and I extended it there to derive the standard expression (first put on a firm foundation by Isaacson 1968) for the local average energy flux in gravitational waves. Feynman was I think right to be disappointed that his argument at Chapel Hill seemed to impress no one and was not taken up and developed by relativists at the time.

I think this example goes to the heart of the question.  Feynman was asking a physicist's question, about how gravitational waves act. All he wanted was a convincing intuitive argument that the waves were real and that he could treat them as part of the rest of physics, for example by extracting thermal energy from them. The relativists of his day, on the other hand, were not interested in this kind of physicist's answer, not even apparently as a first step toward a more complete understanding of gravitational waves. Instead, they seemed to want to transplant as much of the apparatus of energy conservation as they could from the rest of classical physics.  Energy conservation is of course a key concept in theoretical physics.  But the work of Emmy Noether had shown long before that one should not expect exact energy conservation in the absence of time invariance, e.g.\ in a space-time containing gravitational waves.  In relativity energy will always be a subsidiary concept, valid in some circumstances and useless in others.  I believe Feynman found it intensely frustrating that relativists seemed more interested in the pure-radiation energy concept -- in other words, relativity for its own sake -- than in exploring the interaction of gravitational waves with material systems -- gravitational waves as part of physics.

Feynman's example was more than just symptomatic of the way mainstream physics reacted to relativity. Feynman was one of the few mainstream physicists who attempted to cross over the gulf in the 1950s, and he was a prominent opinion-former. Relativity might have been accepted back into the mainstream physics community much earlier if relativists had succeeded in establishing a fruitful dialog with Feynman. Instead, his well-publicized scorn surely damaged the standing of the relativity community materially.

\bigskip\noindent{\large\bf 5. Einstein and the Gulf}\\[2mm]
It is hard to escape the conclusion that Einstein himself was one of the main reasons that the relativity community found itself excluded from mainstream physics.  His influence on relativity research was naturally enormous.  He appears to have rejected the idea of gravitational collapse, for reasons that today are hard to understand.  He also appears not to have been comfortable with gravitational waves, troubled by the coordinate problems. Coordinates were a particular issue, as Silvio Bergia emphasized at this meeting in connection with the issue of general covariance, a principle that seems to have inhibited the development of heuristic concepts until Wheeler began emphasizing a more explicitly geometrical perspective on gravity. Perhaps most importantly, Einstein was focused mainly on finding a unified field theory. He does not seem to have been interested in the importance that general relativity had in classical theoretical physics, still less in its potential in astronomy. Einstein's key bridge to mainstream physics was the unified field theory. Its failure seems to have left relativity without any other bridges.

One further aspect of Einstein's position that I believe may have been important was his rejection of the standard interpretation of quantum mechanics. Naturally, this isolated him from mainstream physics thinking. Perhaps Bohr's advice to Chandrasekhar not to go into relativity had at least something to do with this. But I think there was a more profound way in which Einstein's rejection of quantum heuristics hurt relativity. As I mentioned earlier, quantum physics changed the philosophy of theoretical physics. The key objective of quantum theory became the observable: don't try to describe or understand something that you cannot measure.  Relativity could have benefited in the period 1930--1950 from this imperative to focus only on what is -- at least in principle --  measurable.

It is paradoxical that quantum physicists focused on the importance of observables long before relativists did. After all, coordinate-invariance was a key tenet of general relativity. The difference between quantum theorists and relativists is that in the quantum field the principle was practiced, while in relativity there seems to have been no systematic effort to focus on measurables as a way to solve coordinate difficulties until the ``revival'' began.

I remember, as a graduate student, my supervisor Kip Thorne emphasizing that if coordinate confusion threatened, then one should construct a thought experiment and worry only about what the experimenter could in principle measure; and he made it clear that his own supervisor, John Archibald Wheeler, had emphasized this to him. Wheeler, of course, had worked extensively on quantum physics before taking up relativity in the mid-1950s. The idea of focussing on observables was natural to him and to other physicists of his generation. It had unfortunately not developed sufficiently in relativity, and it seems clear to me that introducing the strict discipline of observability was essential to ending relativity's isolation.

\bigskip\noindent{\large\bf 6. The Gulf Closes}\\[2mm]
It is arguable that a key reason that relativity pulled out of its doldrums was that new blood entered the field with this maxim from quantum theory deeply ingrained in their physical thinking.  For people like Bondi, Pascual Jordan, Wheeler, and Yakov Zel'dovich, among others, it was natural to test any question about general relativity with the demand that it be phrased in terms of observables.  Is there something you can measure, at least an experiment in principle?

At a stroke this way of thinking forces you, for example, to look for other physical features of the black-hole horizon than just the bad behavior of some metric components. Do the local tidal stretching forces near the Schwarzschild ``singularity'' remain finite? Can a real body reach and cross this surface in a finite amount of its own time?

Regarding gravitational waves, this perspective leads you to ask whether a radiating body experiences a back reaction that changes its observable behavior, and whether the radiated gravitational waves in turn produce a measurable effect in the detector.  It is then natural to ask under what circumstances it is reasonable to expect that a definition of energy exists that plays a role in self-gravitating systems analogous to what physicists are used to in nonrelativistic physics; but the energy question does not stop you from answering the questions about observable physical effects of gravitational waves.

It may unfortunately not be a coincidence that relativity began climbing out of the doldrums at about the same time that Einstein died.  His disappearance left the subject open for people to come in who had a background in mainstream physics and who were asking different kinds of questions.  A large number of people working actively in classical theoretical relativity today (leaving aside the quantum gravity and string theory communities) can trace their lineage back to a handful of key physicists who entered relativity between about 1950 and 1960. These physicists reinvigorated the subject by asking the right kinds of questions, and they answered these questions with new heuristic notions that enabled relativity to communicate with and fit into the rest of physics.

Nothing illustrates this change better than the evolution of the black hole concept, to which I referred earlier. The term ``black hole'' was coined as late as 1967 by Wheeler to describe something whose reality he initially also doubted, but which he finally came to understand was the likely endpoint for the evolution of a large range of massive systems.  Today we talk about black holes, not just the Kerr metric or the Schwarzschild solution. That is because the concept of a black hole is wider than just these time-independent exact solutions of Einstein's vacuum field equations. Wheeler himself took a major step toward our present picture by showing, with Tullio Regge, that the Schwarzschild horizon and exterior are stable against small perturbations. Immediately this meant that the idealized Schwarzschild solution was robust enough to include in models of more complicated physical systems: it would retain its essential properties even when disturbed. This robust object is what we call the black hole.

Once the new generation of mathematical physicists recognized that their job was to develop a heuristic understanding of this object, they set to work. An immense number of research papers between 1960 and 1990, including some remarkably elegant mathematics, led to the modern concept of a black hole.

This concept is far wider than the exact Kerr solution of Einstein's equations. Black holes can have accretion disks around them, in which case they are not Kerr. They can have matter falling into them, so they need not even be time-independent. They can convert matter into energy, as Penrose showed. They radiate thermal radiation, as Stephen Hawking showed. They even obey the laws of thermodynamics.

When an astronomer and a relativist talk about black holes, they need this common heuristic concept. In order to use black holes in models for astronomical systems, the astronomer needs to regard the black hole as a kind of black box, an object whose inputs and outputs are known but whose inner workings can be ignored. The astronomer wants to feel safe that he can put a black hole into a binary system without worrying about the details of the horizon or the curvature singularity inside. Relativists today are able to provide astronomers with this black-hole black box.

\bigskip\noindent{\large\bf 7. General Relativity Is Part of Physics}\\[2mm]
This is absolutely typical of physical thinking in other fields. Astronomers talk about stars, by which they mean a synthesis of a huge amount of physics. Nobody can even write down the complete mathematics needed to give an adequate description of a star. Nevertheless an astrophysicist knows pretty well what a star is. The same could be said about a laser, a superconductor, the plasma in a tokamak, or even about relatively simple composite systems like atoms, protons, neutrons. Even in front-line research, where such concepts are not settled, physicists work hard to develop them. The string theory community uses very visual and geometrical heuristics to describe their work. The extension of strings to multi-dimensional branes has opened up a rich source of possible phenomenology, and it is striking to me that, when I listen to talks given by theorists about the applications of brane theory to cosmology and to gravitation theory, the speakers often skip completely over the mathematics in favor of drawings that condense the mathematics into visual relationships.

I believe that this is a basic aspect of the way physicists think about physics.  The mathematical representation of the laws of physics is their foundation, but physicists would generally be paralyzed if they could not package up physical systems into heuristic black boxes, confident that they know (or at least someone knows!) enough about their internal complexity to understand how they will interact with each other.

General relativity has a reasonably well-developed set of physical constructs today. This was the reason, for example, that Ted Newman could give his talk without showing any equations: when he talked about black holes and gravitational waves, we all knew what he meant. Or at least, those of you who are not specialists in general relativity knew something about what he meant, and you had faith that those of us who are specialists knew sufficiently more about what he meant for it to be safe for us all to talk about these concepts as physical reality! Without that faith, physics would simply not be possible. In the 1930s relativity had few such heuristic concepts to offer, and it did not look like it was moving toward constructing many more of them. I suggest that this is what led to the big gulf between relativists and mainstream theoretical physicists between 1930 and 1950. If this picture is right, then general relativity emerged mathematically complete in 1916, but as a theory of physics it was not completed until the 1980s. This must be one of the most gradual of Kuhnian revolutions ever!

\newpage
\bigskip\noindent{\large\bf References}\\[2mm]

\noindent DeWitt, CŽcile M.,  and Rickles, Dean (eds.), (2011) {\em The Role of Gravitation in Physics: Report from the 1957 Chapel Hill Conference}, Max Planck Research Library for the History and Development of Knowledge (Sources 5) (Max Planck Institute for the History of Science, Berlin, Germany).\\
Eisenstaedt, Jean (1989), ``The low water mark of general relativity, 1925-1955'', in {\em Einstein and the History of General Relativity}, eds.\ Howard, Don, and Stachel, John, Birkh\"auser,  Boston, MA, 277-292.\\
Eisenstaedt, Jean (2006), {\em The Curious History of Relativity : how Einstein's Theory of Gravity was Lost and Found Again}, Princeton University Press, Princeton.\\
Feynman, Richard P.\ (1988), {\em What Do You Care What Other People Think?}, W.W.\ Norton, New York.\\
Infeld, Leopold (1964), {\em Relativistic Theories of Gravitation}, Pergamon Press, Oxford.\\
Isaacson, Richard (1968), ``Gravitational Radiation in the Limit of High Frequency. II. Nonlinear Terms and the Effective Stress Tensor'', {\em Physical Review} {\bf 166}, 1272--1280. \\
Robinson, Ivor, Schild, Alfred, and Sch\"ucking, Engelbert L.\ (1965), {\em Quasi-stellar sources and gravitational collapse}, University of Chicago Press, Chicago.\\
Schutz, Bernard F.\ (2009), {\em A First Course in General Relativity} (2nd ed.), Cambridge University Press, Cambridge.\\
Thorne, Kip S.\ (1994), {\em Black Holes and Time Warps}, W.W.\ Norton, New York.

\end{document}